\documentclass[12pt]{article}
\usepackage{latexsym}
\usepackage{amsmath}
\usepackage{amssymb}
\usepackage{amsfonts}
\usepackage{comment}
\usepackage{graphicx}
\usepackage{hyperref}
\usepackage{verbatim}
\usepackage{enumerate,bbold}
\newcommand{\be}{\begin{equation}}
\newcommand{\ee}{\end{equation}}
\newcommand{\bea}{\begin{eqnarray}}
\newcommand{\eea}{\end{eqnarray}}

\newcommand{\calh}{{\mathcal H}}

\newcommand{\htilde}{\widetilde{H}}

\begin{document}
\baselineskip=20pt

\begin{flushright}
CALT-TH-2014-141  \end{flushright}

\begin{center}
{\Large{\bf In What Sense Is the Early Universe Fine-Tuned?}}

\vspace*{0.2in}
Sean M.\ Carroll

\it Walter Burke Institute for Theoretical Physics, California Institute of Technology\\ {\tt seancarroll@gmail.com}
\vspace*{0.1in}
\end{center}

\begin{abstract}
It is commonplace in discussions of modern cosmology to assert that the early universe began in a  special state.
Conventionally, cosmologists characterize this fine-tuning in terms of the horizon and flatness problems.
I argue that the fine-tuning is real, but these problems aren't the best way to think about it: causal disconnection of separated regions isn't the real problem, and flatness isn't a problem at all.
Fine-tuning is better understood in terms of a measure on the space of trajectories: given reasonable conditions in the late universe, the fraction of cosmological histories that were smooth at early times is incredibly tiny.
This discussion helps clarify what is required by a complete theory of cosmological initial conditions.
(Prepared for a volume of essays commemorating David Albert's {\it Time and Chance}, B.\ Loewer, E.\ Winsberg and B.\ Weslake, eds.)
\end{abstract}
%\vfill\newpage
%\newpage

\baselineskip=14pt
\tableofcontents

%%%%%%%%%%%%%%%%%%%%%%%%%%%
\section{Introduction}\label{introduction}

The issue of the initial conditions of the universe -- in particular, the degree to which they are ``unnatural'' or ``fine-tuned,'' and possible explanations thereof -- is obviously of central importance to cosmology, as well as to the foundations of statistical mechanics. 
The early universe was a hot, dense, rapidly-expanding plasma, spatially flat and nearly homogeneous along appropriately chosen spacelike surfaces.%
\footnote{Our concern here is the state of the universe -- its specific configuration of matter and energy, and the evolution of that configuration through time -- rather than the coupling constants of our local laws of physics, which may also be fine-tuned.
I won't be discussing the value of the cosmological constant, or the ratio of dark matter to ordinary matter, or the matter/antimatter asymmetry.} 
The question is, \emph{why} was it like that?
In particular, the thinking goes, these conditions don't seem to be what we would expect a ``randomly chosen'' universe to look like, to the extent that such a concept makes any sense.
In addition to the obvious challenge to physics and cosmology of developing a theory of initial conditions under which these properties might seem natural, it is a useful exercise to specify as carefully as possible the sense in which they don't seem natural from our current point of view.

Philosophers of science (and some physicists \cite{Penrose,Carroll:2004pn}) typically characterize the kind of fine-tuning exhibited by the early universe as being a state of \emph{low entropy}.
This formulation has been succinctly captured in David Albert's {\it Time and Chance} as ``The Past Hypothesis'' \cite{albert}.
A precise statement of the best version of the Past Hypothesis is the subject of an ongoing discussion (see {\it e.g.} \cite{earman2006past,wallace2011}). 
But it corresponds roughly to the idea that the early universe -- at least the observable part of it, the aftermath of the hot Big Bang -- was in a low-entropy state with the right micro-structure to evolve in a thermodynamically sensible way into the universe we see today.

Cosmologists, following Alan Guth's influential paper on the inflationary-universe scenario \cite{Guth:1980zm}, tend to describe the fine-tuning of the early universe in terms of the horizon and flatness problems.\footnote{Guth also discussed the overabundance of magnetic monopoles predicted by certain grand unified theories. This problem was the primary initial motivation for inflation, but is model-dependent in a way that the horizon and flatness problems don't seem to be.}
The horizon problem, which goes back to Misner \cite{misner1969mixmaster}, is based on the causal structure of an approximately homogeneous and isotropic (Friedmann-Robertson-Walker) cosmological model in general relativity.
If matter and radiation (but not a cosmological constant) are the only sources of energy density in the universe, regions we observe using the cosmic background radiation that are separated by more than about one degree were never in causal contact -- their past light-cones, traced back to the Big Bang, do not intersect.
It is therefore mysterious how they could be at the same temperature, despite the impossibility in matter/radiation cosmology of any signal passing from one such point to another.
The flatness problem, meanwhile, was elucidated by Dicke and Peebles \cite{dicke1979big}. 
Spatial curvature grows with respect to the energy density in matter or radiation, so it needs to be extremely small at early times so as not to be completely dominant today.
(For some history of the horizon and flatness problems, see \cite{brawer1995inflationary}.
Brawer notes that neither the horizon problem nor the flatness problem were considered of central importance to cosmology until inflation suggested a solution to them.)

The horizon and flatness problems and the low entropy of the early universe are clearly related in some way, but also seem importantly different.
In this essay I will try to clarify the nature of the horizon and flatness problems, and argue that they are \emph{not} the best way of thinking about the fine-tuning of the early universe.

The horizon problem gestures in the direction of a real puzzle, but the actual puzzle is best characterized as fine-tuning within the space of cosmological \emph{trajectories}, rather than an inability of the early universe to equilibrate over large distances.
This reformulation is important for the status of inflationary cosmology, as it makes clear that inflation by itself does not solve the problem (though it may play a crucial role in an ultimate solution).
The flatness problem, meanwhile, turns out to be simply a misunderstanding; the correct measure on cosmological trajectories predicts that all but a set of measure zero should be spatially flat.
Correctly describing the sense in which the early universe is fine-tuned helps us understand what kind of cosmological models physicists and philosophers should be endeavoring to construct.

%%%%%%%%%%%%%%%%%%%%%%%%%%%%%%%%%%%%%%%%%%%%%
\section{What Needs to be Explained}\label{what}

In order to understand the claim that the state of the universe appears fine-tuned, we should specify what features of that state we are talking about.
According to the standard cosmological model, the part of the universe we are able to observe is expanding, and emerged from a hot, dense Big Bang about fourteen billion years ago.%
\footnote{In classical general relativity such a state corresponds to a curvature singularity; in a more realistic but less well-defined quantum theory of gravity, what we call the Big Bang may or may not have been the absolute beginning of the universe, but at the least it is a moment prior to which we have no empirical access.}
The distribution of matter and radiation at early times was nearly uniform, and the spatial geometry was very close to flat.
We see the aftermath of that period in the cosmic microwave background (CMB), radiation from the time the universe became transparent about 380,000 years after the Big Bang, known as the ``surface of last scattering.''
This radiation is extremely isotropic; its observed temperature is 2.73~K, and is smooth to within about one part in $10^5$ across the sky \cite{Ade:2013ktc}.
Our observable region contains about $10^{88}$ particles, most of which are photons and neutrinos, with about $10^{79}$ protons, neutrons, and electrons, as well as an unknown number of dark matter particles.
(The matter density of dark matter is well-determined, but the mass per particle is extremely uncertain.)
The universe has evolved today to a collection of galaxies and clusters within a web of dark matter, spread homogeneously on large scales over an expanse of tens of billions of light years.

Note that our confidence in this picture depends on assuming the Past Hypothesis.
The most relevant empirical information concerning the smoothness of the early universe comes from the isotropy of the CMB, but that does not strictly imply uniformity at early times.
It is a geometric fact that a manifold that is isotropic around every point is also homogeneous; hence, the observation of isotropy and the assumption that we do not live in a special place together are often taken to imply that the matter distribution on spacelike slices is smooth.
But we don't observe the surface of last scattering directly; what we observe is the isotropy of the temperature of the radiation field reaching our telescopes here in a much older universe.
That temperature is determined by a combination of two factors: the intrinsic temperature at emission, and the subsequent redshift of the photons.
The intrinsic temperature at the surface of last scattering is mostly set by atomic physics, but not quite; because there are so many more photons than atoms, recombination occurs at a temperature of about 3\,eV, rather than the 13.6\,eV characterizing the ionization energy of hydrogen. 
Taking the photon-to-baryon ratio as fixed for convenience, our observations therefore imply that the cosmological redshift is approximately uniform between us and the surface of last scattering in all directions on the sky.
But that is compatible with a wide variety of early conditions.
The redshift along any particular path is a single number; it is not possible to decompose it into ``Doppler'' and ``cosmic expansion'' terms in a unique way \cite{bunn2009}.
This reflects the fact that there are many ways to define spacelike slices in a perturbed cosmological spacetime.
For example, we can choose to define spacelike surfaces such that the cosmological fluid is at rest at every point (``synchronous gauge'').
In those coordinates there is no Doppler shift, by construction.
But the matter distribution at recombination could conceivably look very inhomogeneous on such slices; that would be compatible with our current observations as long as a direction-dependent cosmological redshift conspired to give an isotropic radiation field near the Earth today.
Alternatively, we could choose spacelike slices along which the temperature was constant; the velocity of the fluid on such slices could be considerably non-uniform, yet the corresponding Doppler effect could be canceled by the intervening expansion of space along each direction.
Such conspiratorial conditions seem unlikely to us, but they are more numerous (in the measure to be discussed below) in the space of all possible initial conditions.
Of course, we also know that most past conditions that lead to a half-melted ice cube in a glass of water look like a glass of liquid water at a uniform temperature, rather than the unmelted ice cube in warmer water we would generally expect.
In both cases, our conventional reasoning assumes the kind of lower-entropy state postulated by the Past Hypothesis.

With the caveat that the Past Hypothesis is necessary, let us assume that the universe we are trying to account for is one that was very nearly uniform (and spatially flat) at early times.
What does it mean to day that such a state is fine-tuned?

Any fine-tuning is necessarily a statement about one's expectations about what would seem natural or non-tuned. 
In the case of the initial state of the universe, one might reasonably suggest that we simply have no right to have any expectations at all, given that we have only observed one universe.
But this is a bit defeatist.
While we have not observed an ensemble of universes from which we might abstract ideas about what a natural one looks like, we know that the universe is a physical system, and we can ask whether there is a sensible \emph{measure} on the relevant space of states for such a system, and then whether our universe seems generic or highly atypical in that measure.
In practice we typically use the classical Liouville measure, as we will discuss in Section~\ref{trajectories}.
The point of such an exercise is not to say ``We have a reliable theory of what universes should look like, and ours doesn't fit."
Rather, given that we admit that there is a lot about physics and cosmology that we don't yet understand, it's to look for clues in the state of the universe that might help guide us toward a more comprehensive theory.
Fine-tuning arguments of this sort are extremely important in modern cosmology, although the actual measure with respect to which such arguments are made are sometimes insinuated rather than expressly stated, and usually posited rather than being carefully derived under some more general set of assumptions.

The fine-tuning argument requires one element in addition to the state and the measure: a way to coarse-grain the space of states.
This aspect is often overlooked in discussions of fine-tuning, but it is again implicit.
Without a coarse-graining, there is no way to say that any particular state is ``natural'' or ``fine-tuned,'' even in a particular measure on the entire space of states.
Each individual state is just a point, with measure zero.
What we really mean to say is that states \emph{like} that state are fine-tuned, in the sense that the corresponding macrostate in some coarse-graining has a small total measure.
The coarse-graining typically corresponds to classifying states as equivalent if they display indistinguishable macroscopically observable properties.
It is usually not problematic, although as we will see it is necessary to be careful when we consider quantities such as the spatial curvature of the universe.

Given a measure and a coarse-graining, it is natural to think of the entropy of each state, defined by Boltzmann to be the logarithm of the volume of the corresponding macrostate.
Penrose has famously characterized the fine-tuning problem in these terms, emphasizing how small the entropy of the early universe was compared to the current entropy, and how small that is compared to how large the entropy could be \cite{Penrose}.
At early times, when inhomogeneities were small, we can take the entropy to simply be that of a gas of particles in the absence of strong self-gravity, which is approximately given by the number of particles; for our observable universe, that's about $10^{88}$.
Today, the entropy is dominated by supermassive black holes at the center of galaxies, each of which has a Bekenstein-Hawking entropy
\be
  S_{\mathrm{BH}} = \frac{A}{4G} = 4\pi G M^2
\ee
in units where $\hbar = c = k_B = 1$.
The best estimates of the current entropy give numbers of order $10^{103}$ \cite{egan2010larger}.
Penrose suggests a greatest lower bound on the allowed entropy of our observable universe by calculating what the entropy would be if all of the matter were put into one gigantic black hole, obtaining the famous number $10^{122}$.
Since the universe is accelerating, distant galaxies will continue to move away from us rather than collapsing together to form a black hole, but we can also calculate the entropy \cite{Banks:2000fe,Banks:2001yp} of the de~Sitter phase to which the universe will ultimately evolve, obtaining the same number $10^{122}$. 
(The equality of these last two numbers can be explained by the cosmological coincidence of the density in matter and vacuum energy; they are both of order $1/GH^2_0$, where $H_0$ is the current Hubble parameter.)
By any measure, the entropy of the early universe was fantastically smaller than its largest possible value, which seems to be a fine-tuning.%
\footnote{Even though the comoving volume corresponding to our observable universe was much smaller at early times, it was still the same physical system as the late universe, with the same space of states, governed by presumably-reversible dynamical laws.
It is therefore legitimate to calculate the maximum entropy of the early universe by calculating the maximum entropy of the late universe.}
Part of our goal in this essay is to relate this formulation to the horizon and flatness problems.

To close this section, we note that the early universe was in an extremely \emph{simple} state, in the sense that its ``apparent complexity'' is low: macroscopic features of the state can be fully characterized by a very brief description (in contrast with the current state, where a macroscopic description would still require specifying every galaxy, if not every star and planet).
But apparent complexity is a very different thing than entropy; the universe is simple at the earliest times, and will be simple again at much later times, while entropy increases monotonically \cite{aaronson}.
The simplicity of the early universe should not in any way be taken as a sign of its ``naturalness.''
A similar point has been emphasized by Price, who notes that any definition of ``natural'' that purportedly applies to the early universe should apply to the late universe as well \cite{Price:1993hr}.
The early universe appears fine-tuned because the macroscopic features of its state represent a very small region of phase space in the Liouville measure, regardless of how simple it may be.

%%%%%%%%%%%%%%%%%%%%%%%%%%%%%%%%%%%%%%%%%%%%%
\section{The Horizon Problem}\label{horizon}

%%%%%%%%%%%%%%%%%%%%%%%%%%%%%%%%%%%%%%%%%%%%%
\subsection{Defining the problem}

We now turn to the horizon problem as traditionally understood.
Consider a homogeneous and isotropic universe with scale factor $a(t)$, energy density $\rho(a)$, and a fixed spatial curvature $\kappa$. In general relativity, the evolution of the scale factor is governed by the Friedmann equation, 
\be
%  H^2 = \frac{1}{3\mpl^2}\rho - \frac{\kappa}{a^2},
  H^2 = \frac{8\pi G}{3}\rho - \frac{\kappa}{a^2},
  \label{friedmann}
\ee
where $H=\dot{a}/a$ is the Hubble parameter and %$\mpl = 1/\sqrt{8\pi G}$ is the reduced Planck mass (in units where $\hbar = c = 1$). 
$G$ is Newton's constant of gravitation, and we use units where the speed of light is set equal to unity, $c=1$.
Often, the energy density will evolve as a simple power of the scale factor, $\rho \propto a^{-n}$. For example, ``matter'' to a cosmologist is any species of massive particles with velocities much less than the speed of light, for which $\rho_M \propto a^{-3}$, while ``radiation'' is any species of massless or relativistic particles, for which $\rho_R \propto a^{-4}$. 
(In both cases, the number density decreases as the volume increases, $n\propto a^{-3}$; for matter the energy per particle is constant, while for radiation it diminishes as $a^{-1}$ due to the cosmological redshift.)

When the energy density of the universe is dominated by components with $n > 2$, $\dot a$ will be decreasing, and we say the universe is ``decelerating.'' 
In a decelerating universe the horizon size at time $t_*$ (the distance a photon can travel between the Big Bang and $t_*$) is approximately
\be
  d_{\mathrm{hor}}(t_*) \approx t_* \approx H_*^{-1}.
\ee
We call $H^{-1}$ the ``Hubble distance'' at any given epoch.
Sometimes the Hubble distance is conflated with the horizon size, but they are importantly different; the Hubble distance depends only on the instantaneous expansion rate at any one moment in time, while the horizon size depends on the entire past history of the universe back to the Big Bang.
The two are of the same order of magnitude in universes dominated by matter and radiation, with the precise numerical factors set by the abundance of each component.
A textbook calculation shows that the Hubble distance at the surface of last scattering, when the CMB was formed, corresponds to approximately one degree on the sky today.

The horizon problem, then, is  simple.
We look at widely-separated parts of the sky, and observe radiation left over from the early universe.
As an empirical matter, the temperatures we observe in different directions are very nearly equal.
But the physical locations from which that radiation has traveled were separated by distances larger than the Hubble distance (and therefore the horizon size, if the universe is dominated by matter and radiation) at that time.
They were never in causal contact; no prior influence could have reached both points by traveling at speeds less than equal to that of light.
Yet these independent regions seem to have conspired to display the same temperature to us.
That seems like an unnatural state of affairs. 
We can formalize this as:

\begin{quote}
{\bf Horizon problem (causal version).} 
If different regions in the early universe have non-overlapping past light cones, no causal influence could have coordinated their conditions and evolution.
There is therefore no reason for them to appear similar to us today.
\end{quote}

If that's as far as it goes, the horizon problem is perfectly well-formulated, if somewhat subjective.
The causal formulation merely points out that there is no reason for a certain state of affairs (equal temperatures of causally disconnected regions) to obtain, but it doesn't give any reason for expecting otherwise.
Characterizing this as a ``problem,'' rather than merely an empirical fact to be noted and filed away, requires some positive expectation for what we think conditions near the Big Bang \emph{should} be like: some reason to think that unequal temperatures would be more likely, or at least less surprising. 

%%%%%%%%%%%%%%%%%%%%%%%%%%%%%%%%%%%%%%%%%%%%%
\subsection{Equilibration and entropy}

We can attempt to beef up the impact of the horizon problem by bringing in the notion of \emph{equilibration} between different regions.
Imagine we have a box of gas containing different components, or simply different densities.
If it starts from an inhomogeneous (low-entropy) configuration, given time the gas will typically equilibrate and attain a uniform distribution.
In that sense, an equal temperature across a hot plasma actually seems natural or likely.
But if we think of the early universe as such a box of gas, it hasn't had time to equilibrate.
That's a direct consequence of the fact that (in a matter/radiation dominated universe) the horizon size is much smaller than the scales over which we are currently observing.
Let's call this the ``equilibration'' version of the horizon problem.
\begin{quote}
{\bf Horizon problem (equilibration version).} 
If different regions in the early universe shared a causal past, they could have equilibrated and come to the same temperature.
But when they do not, such as in a matter/radiation-dominated universe, equal temperatures are puzzling.
\end{quote}

The equilibration formulation of the horizon problem seems stronger than the causal version; it attempts to provide some reason why equal temperatures across causally disconnected reasons should be surprising, rather than merely noting their existence.
But in fact this version undermines the usual conclusions that the horizon problem is intended to justify, and invites a critique that has been advanced by Sheldon Goldstein \cite{goldsteinbox}.
I will try to argue that Goldstein's critique doesn't rebut the claim that the early universe is fine-tuned, though it does highlight the misleading nature of the equilibration version of the horizon problem.

The problem arises when we try to be more specific about what might count as ``natural'' initial conditions in the first place.
If we trace the Friedmann equation (\ref{friedmann}) backwards in time, we come to a singularity -- a point where the scale factor is zero and the density and Hubble parameter are infinite.
There is no reason to expect the equations of classical general relativity to apply in such circumstances; at the very least, quantum effects will become important, and we require some understanding of quantum gravity to make sensible statements. 
Absent such a theory, there would be some justification in giving up on the problem entirely, and simply saying that the issue of initial conditions can't be addressed without a working model of quantum gravity.
(Presumably this was the feeling among many cosmologists before inflation was proposed.)

Alternatively, we could choose to be a bit more optimistic, and ask what kind of configurations would constitute natural initial conditions in the moments right after the initial singularity or whatever quantum phase replaces it.
Given the phase space describing the relevant degrees of freedom in that regime, we can coarse-grain into macrostates defined by approximately equal values of macroscopically observable quantities and define the Boltzmann entropy as the logarithm of the volume of the macrostate of which the microstate is an element, as discussed in Section~\ref{what}.
Calculating this volume clearly requires the use of a measure on the space of states; fortunately such a measure exists, given by Liouville in the classical case or by the Hilbert-space inner product in the quantum case.
Given that machinery, states with high Boltzmann entropy seem natural or generic or likely, simply because there are many of them; states with low Boltzmann entropy seem unnatural since they are relatively few, and suggest the need for some sort of explanation.%
\footnote{Note that the expectation of high entropy for the early universe, whether convincing or not, certainly has a different character than our expectation that long-lived closed systems will be high-entropy in the real world.
In the latter case, there is a dynamical mechanism at work: almost all initial conditions will evolve toward equilibrium.
In the case of the early universe, by contrast, we are actually making a statement about the initial conditions themselves, saying that high-entropy ones would be less surprising than low-entropy ones, since the former are more numerous.}

Let's apply this to a matter/radiation-dominated universe, in which different regions we observe in the CMB are out of causal contact.
Following Goldstein, we can draw an analogy with two isolated boxes of gas.
The boxes could be different sizes, and have never interacted.
All we know is that there is some fixed number of particles in the boxes, with some fixed energy.
Goldstein's observation is that, if we know nothing at all about the particles inside the two boxes, it should be \emph{completely unsurprising} if they had the same temperature.
The reasoning is simple: at fixed energy and particle number, there is more phase space corresponding to approximately equal-temperature configurations than to unequal-temperature ones.
Such configurations maximize the Boltzmann entropy.
Given the two boxes, some fixed number of particles, and some fixed total energy, a randomly-chosen point in phase space is very likely to feature equal temperatures in each box, even if they have never interacted.
Therefore, one might tentatively suggest, perhaps seeing equal temperatures in causally disconnected parts of the early universe isn't actually unlikely at all, and the horizon problem shouldn't be a big worry.

%%%%%%%%%%%%%%%%%%%%%%%%%%%%%%%%%%%%%%%%%%%%%
\subsection{Gravity and dynamics}

For isolated boxes of gas, this logic is surprising, but valid.
A randomly-selected state of two isolated boxes of gas is likely to have equal temperatures in each box, even in the absence of interactions.
For the early universe, however, the boxes turn out to not provide a very useful analogy, for a couple of (related) reasons: the importance of gravity, and the fact that we are considering time-dependent trajectories, not simply individual states.

Gravity plays the crucial role in explaining a peculiar feature of the early universe: it is purportedly a low-entropy state, but one in which the matter is radiating as a nearly-homogeneous black body, exactly as we are trained to expect from systems in thermal equilibrium.
The simple resolution is that, when the \emph{self-gravity} of a collection of particles becomes important, high-entropy configurations of specified volume and density are inhomogeneous rather than homogeneous.
Therefore, if the imaginary boxes that we consider are sufficiently large, we wouldn't expect the temperature to be uniform even inside a single box, much less between two boxes.

This can be seen in a couple of different ways. 
One is to consider the Jeans instability: the tendency of sufficiently long-wavelength perturbations in a self-gravitating fluid to grow.
In a fluid with density $\rho$ and speed of sound $c_s$, modes are unstable to growth if they are larger than the Jeans length,
\be
 \lambda_J = \frac{c_s}{(G\rho)^{1/2}}.
\ee
If the size of a box of gas is greater than the Jeans length for that gas, the distribution will fragment into self-gravitating systems of size $\lambda_J$ or smaller, rather than remaining homogeneous; in cosmology, that's the process of structure formation.
Given the fluid pressure $p$ as a function of the energy density, the speed of sound is defined by $c_s^2 = dp/d\rho$.
For a radiation bath it is $c_R = 1/\sqrt{3}$, while for collisionless nonrelativistic particles we have $c_M\approx 0$; a cosmological matter fluid is unstable to growth of inhomogeneities on all scales.%
\footnote{This story is complicated by the expansion of the universe, since the Hubble parameter acts as a friction term; in a purely radiation-dominated Friedmann-Robertson-Walker universe, density perturbations actually shrink as the universe expands.
But even when radiation dominates the energy density, there are still slowly-moving matter particles.
Given sufficient time in a hypothetical non-expanding universe, density perturbations in non-relativistic matter would grow very large.
This leads us to the discussion of trajectories rather than states, which we undertake in the next section.}
Classically, the only perfectly stable configuration of fixed energy in a fixed volume would be a black hole in vacuum, which is about as inhomogeneous as you can get.

Another way of reaching the same conclusion (randomly-chosen states of self-gravitating particles are likely to be inhomogeneous) is to examine phase-space volumes directly.
Consider a collection of particles with a given energy, interacting only through the inverse-square law of Newtonian gravity.
It is clear that the volume of phase space accessible to such a system is unbounded.
Keeping the energy fixed, we can send any number of particles to infinity (or arbitrarily large momentum) while compensating by moving other particles very close together, sending their mutual energy to minus infinity.
This is a real effect, familiar to researchers in galactic dynamics as the ``gravo-thermal catastrophe'' \cite{lynden-bell,nityananda2009gravitational}.
A galaxy populated by stars (or dark matter particles) interacting through Newtonian gravity will tend to become centrally condensed without limit, while ejecting other stars to infinity.
The entropy of such a system is unbounded, and there is no equilibrium configuration, but generic evolution is in the direction of greater inhomogeneity.
And, indeed, in semiclassical general relativity, the highest-entropy configuration of fixed energy in a fixed volume is generally a black hole when the energy is sufficiently high.%
\footnote{When the energy is too low, any black hole will have a higher Hawking temperature than its surroundings, and it will lose energy and shrink. 
Its temperature will grow as the hole loses mass, and eventually it will evaporate completely away.
This is a reflection of the fact that black holes have negative specific heat.}

In the early universe, the Jeans length is generally less than the Hubble radius (although they are of the same order for purely-radiation fluids). 
High-entropy states will generally be very inhomogeneous, in stark contrast with the intuition we have from boxes of gas with negligible self-gravity, in accordance with Penrose's analysis mention in Section~\ref{what}.
Hence, the equilibration version of the horizon problem is extremely misleading, if not outright incorrect.
In an alternative world in which particles could still gravitationally attract each other but the universe was not expanding, so that past light cones stretched forever and the horizons of any two points necessarily overlapped, our expectation would \emph{not} be for a smooth universe with nearly constant temperatures throughout space.
It would be for the opposite: a highly inhomogeneous configuration with wildly non-constant temperatures.
Whatever the fine-tuning problem associated with the early universe may be, it is not that ``distant regions have not had time to come into thermal equilibrium."
Indeed, the lack of time to equilibrate is seen to be a feature, rather than a bug: it would be even harder to understand the observed uniformity of the CMB if the plasma had had an arbitrarily long time to equilibrate.

From this perspective, the thermal nature of the CMB radiation is especially puzzling.
It cannot be attributed to ``thermal equilibrium,'' since the early plasma is not in equilibrium in any sense.%
\footnote{This is trivial, of course. Equilibrium states are time-independent, while the early universe is expanding and evolving. More formally, there is no timelike Killing vector.}
Sometimes an attempt is made to distinguish between ``gravitational" and ``non-gravitational'' degrees of freedom, and argue that the non-gravitational degrees of freedom are in equilibrium, while the gravitational ones are not. 
This is problematic at best.
The relevant gravitational effect isn't one of degrees of freedom (which would be independently-propagating gravitational waves or gravitons), but the simple existence of gravity as a force.
One might try to argue that the primordial plasma looks (almost) like it would look in an equilibrium configuration in a universe where there was no force due to gravity, but it is unclear what force such an observation is supposed to have.%
\footnote{There has been at least one attempt to formalize this notion, by inventing a scenario in which the strength of gravity goes to zero in the very early universe \cite{Greene:2009tt}.
While intriguing, from the trajectory-centered point of view advocated in the next section even this model doesn't explain why our observed universe exhibits such an unlikely cosmological history.}

The same conclusion can be reached in a complementary way, by recalling that we are considering a highly time-dependent situation, rather than two boxes of gas at rest.
As mentioned in Section~\ref{what}, there is a great amount of freedom in choosing spatial hypersurfaces in a perturbed cosmological spacetime; for example, we could choose the temperature of the fluid as our time coordinate (as long as the fluid was not so inhomogeneous that constant-temperature surfaces were no longer spacelike).
Then, by construction, the temperature is completely uniform at any moment of time.
But we are not observing the CMB at the moment of recombination; the photons we see have been redshifted by a factor of over a thousand.
From that perspective, the question is not ``Why was the temperature uniform on spacelike hypersurfaces?", but rather ``Why is the redshift approximately equal in completely separate directions on the sky?"%
\footnote{Alternatively and equivalently, we could define spacelike hypersurfaces by demanding that each hypersurface be at a constant redshift factor from the present time.
On such surfaces the temperature would generically be very non-uniform.
In these coordinates the question becomes, ``Why did distant regions of the universe reach the recombination temperature at the same time?"}
This formulation suggests the importance of considering cosmological trajectories, which we turn to in Section~\ref{trajectories}.

%%%%%%%%%%%%%%%%%%%%%%%%%%%%%%%%%%%%%%%%%%%%%
\subsection{Inflation}

It is useful to see how inflation addresses the horizon problem.
At the level of the causal version, all inflation needs to do is invoke an extended period of accelerated expansion.
Then the past horizon size associated with a spacetime event becomes much larger than the Hubble radius at that time, and widely-separated points on the surface of last scattering can easily be in causal contact. 
We can quantify the total amount of inflation in terms of the number of $e$-folds of expansion,
\begin{equation}
N_e=\int_{a_{{\rm i}}}^{a_{{\rm f}}}d\ln{a}=\int_{t_{{\rm i}}}^{t_{{\rm f}}}H\, dt.
\end{equation}
where the integral extends from the beginning to the end of the period of accelerated expansion ($\ddot a > 0$).
Generally, at least $N_e > 50$ $e$-folds of inflation are required to ensure that all the observed regions of the CMB share a causal patch.

Whether or not these regions have ``equilibrated'' depends on one's definitions.
During inflation, the energy density of the universe is dominated by the approximately-constant potential energy of a slowly-rolling scalar field.
The evolution is approximately adiabatic, until reheating when inflation ends and the inflaton energy converts into matter and radiation.
Perturbations generically shrink during the inflationary period, in accordance with intuition from the cosmic no-hair theorem \cite{Wald:1983ky}, which states that a universe dominated by a positive cosmological constant approach a de~Sitter geometry.
But it is not a matter of interactions between degrees of freedom in different regions sharing energy, as in a conventional equilibration process; perturbations simply decrease locally and independently as the universe expands.

Once accelerated expansion occurs, it is important that inflation \emph{end} in such a way that homogeneous and isotropic spatial slices persist after reheating (when energy in the inflaton field is converted into matter and radiation).
The success of this step is highly non-trivial; indeed, this ``graceful-exit problem'' was highlighted by Guth \cite{Guth:1980zm} in his original paper, which relied on bubble nucleation to enact the transition from a metastable false vacuum state to the true vacuum.
Unfortunately, there are only two regimes for such a process; if bubbles are produced rapidly, inflation quickly ends and does not last for sufficient $e$-folds to address the horizon problem, while if they are produced slowly, we are left with a wildly inhomogeneous universe where a few bubbles have appeared and collided while inflation continues in the false vacuum elsewhere.
An attractive solution to this dilemma came in the form of slow-roll inflation \cite{Linde:1981mu,Albrecht:1982wi}. 
Here the field is not trapped in a false vacuum, but rolls gradually down its potential.
Slow-roll inflation can very naturally produce homogeneous and isotropic spatial slices, even if it proceeds for a very large number of $e$-folds.
Central to the success of this model is the fact that the rolling field acts as a ``clock,'' allowing regions that have been inflated to extreme distances to undergo reheating at compatible times \cite{Anninos:1991ma}.
It is thus crucially important that the universe during slow-roll inflation is \emph{not} truly in equilibrium, even though its evolution is is approximately adiabatic; the evolving inflaton field allows for apparent coordination among widely-separated regions.

Inflation, therefore, solves the puzzle raised by the horizon problem, in the following sense: given a sufficient amount of inflation, and a model that gracefully exits from the inflationary phase, we can obtain a homogeneous and isotropic universe in which distant points share a common causal past.
The success of this picture can obscure an important point: the conditions required to get inflation started in the first place are extremely fine-tuned.
This fine-tuning is often expressed in terms of entropy; a patch of spacetime ready to begin inflating has an enormously lower entropy than the (still quite low-entropy) homogeneous plasma into which it evolves \cite{Penrose,Carroll:2004pn}.
In this sense, inflation ``explains'' the fine-tuned nature of the early universe by positing an initial condition that is even more fine-tuned.
In the context of the horizon problem, this issue can be sharpened.
One can show that, in order for inflation to begin, we must not only have very particular initial conditions in which potential energy dominates over kinetic and gradient energies, but that the size of the patch over which these conditions obtain must be strictly larger than the Hubble radius \cite{Vachaspati:1998dy}.
In other words, even in an inflationary scenario, it is necessary to invoke smooth initial conditions over super-horizon-sized distances.
As compelling as inflation may be, it still requires some understanding of pre-inflationary conditions to qualify as a successful model.

Contemporary discussion of inflation often side-steps the problem of the required low-entropy initial condition by appealing to the phenomenon of eternal inflation: in many models, if inflation begins at all it continues without end in some regions of the universe, creating an infinite amount of spacetime volume \cite{Guth:2000ka}. 
While plausible (although for some recent concerns see \cite{Boddy:2014eba}), this scenario raises a new problem: rather than uniquely predicting a universe like the kind we see, inflation predicts an infinite variety of universes, making prediction a difficult problem. I won't discuss this issue here, but for recent commentary see \cite{Ijjas:2013vea,Guth:2013sya,Linde:2014nna,Ijjas:2014nta}.

%%%%%%%%%%%%%%%%%%%%%%%%%%%%%%%%%%%%%%%%%%%%%
\section{Trajectories}\label{trajectories}

%%%%%%%%%%%%%%%%%%%%%%%%%%%%%%%%%%%%%%%%%%%%%
\subsection{Initial Conditions and Histories}

The horizon problem, as discussed in the previous section, might not seem like an extremely pressing issue.
The causal version is fairly weak, merely noting a state of affairs rather than offering any reason why we should expect the contrary, while the equilibration version is misleading -- the problem is not that distant regions are unable to equilibrate, it's that equilibration would have made things more inhomogeneous.
Nevertheless, there is no question that the early universe \emph{is} fine-tuned.
A better statement of the fine-tuning problem comes from considering cosmological trajectories -- histories of the universe over time -- rather than concentrating on initial conditions.

The relationship between initial conditions and trajectories is somewhat different in classical mechanics and quantum mechanics.
In classical mechanics, the space of trajectories of a system with a fixed phase space and time-independent Hamiltonian is isomorphic to the space of initial conditions, or indeed the space of possible conditions at any specified moment of time.
We can think of the history of a classical system as a path through a phase space $\Gamma$ with coordinates $\{q^i, p_i\}$, where the $\{q^i\}$ are canonical coordinates and the $\{p_i\}$ are their conjugate momenta, governed by a Hamiltonian $\calh(q^i, p_i)$ (which for our purposes is taken to be time-independent).
Evolution is unitary (information-conserving), implying that the state at any one time plus the Hamiltonian is sufficient to determine the state at any prior or subsequent time.
Classically, however, trajectories (and time itself) can come to an end; that's what happens at spacelike singularities such as the Big Bang or inside a Schwarzschild black hole.
This reflects the great amount of freedom that exists in choosing the geometry of phase space, including the possible existence of singular points on the manifold.
It is therefore possible for some conditions to be truly ``initial,'' if they occur at the past boundary of the time evolution.

In quantum mechanics, there are no singularities or special points in the state space.
A general solution to the Schr\"odinger equation $\hat H |\psi\rangle = i\partial_t|\psi\rangle$ with a time-independent Hamiltonian can be written in terms of energy eigenstates as 
\be 
  |\psi(t)\rangle = \sum_n r_n e^{i(\theta_i - E_nt)}|E_n\rangle,
\ee  
where the constant real parameters $\{r_i, \theta_i\}$ define the specific state.
Each eigenstate merely rotates by a phase, singularities cannot arise, and time extends infinitely toward the past and future \cite{Carroll:2008yd}.%
\footnote{The Borde-Guth-Vilenkin (BGV) theorem \cite{Borde:2001nh} demonstrates that spacetimes with an average expansion rate greater than zero must be geodesically incomplete in the past (which is almost, but not quite, equivalent to saying there are singularities).
This has been put forward as evidence that the universe must have had a beginning \cite{Mithani:2012ii}; there are explicit counterexamples to this claim \cite{Aguirre:2001ks}, but such examples are arguably unstable or at least non-generic.
However, while the BGV theorem does not assume Einstein's equation or any other equations of motion, it only makes statements about classical spacetime.
It is therefore silent on the question of what happens when gravity is quantized.}
In contrast to the arbitrariness of classical phase space, the geometry of pure quantum states is fixed to be ${\mathbb C}P^n$, and evolution is always smooth \cite{kibble1979geometrization,Brody:1999cw,Bengtsson:2001yd}.
There is then no special ``initial'' condition; the state at any one time can be evolved infinitely far into the past and future.
When we come to quantum gravity, canonical quantization of general relativity suggests \cite{Wiltshire:1995vk} that the wave function of the universe may be an exact energy eigenstate with zero eigenvalue, as reflected in the Wheeler-DeWitt equation:
\be
  \hat{H}|\psi\rangle = 0,
\ee
where the Hamiltonian $\hat H$ includes both gravitational and matter dynamics.
In that case there is no time evolution in the conventional sense, although time can conceivably be recovered as an effective concept describing correlations between a ``clock'' subsystem and the rest of the quantum state \cite{Page:1983uc,Banks:1984cw}.
It is nevertheless far from clear that the Wheeler-DeWitt equation is the proper approach to quantum gravity, or indeed that local spacetime curvature is the proper thing to quantize.
Evidence from the holographic principle \cite{Hooft:1993gx,Susskind:1994vu,Bousso:2002ju}, black hole complementarity \cite{Susskind:1993if}, the gauge-gravity correspondence \cite{Maldacena:1997re,Horowitz:2006ct}, the entanglement/spacetime connection \cite{Ryu:2006bv,swingle2009,VanRaamsdonk:2010pw,Maldacena:2013xja}, and thermodynamic approaches to gravity \cite{Jacobson:1995ab,Verlinde:2010hp} suggests that gravity can be thought of emerging from  non-local degrees of freedom, only indirectly related to curved spacetime.
Given the current state of the art, then, it is safest to leave open the question of whether time is emergent or fundamental, and whether it is eternal or has a beginning.

Fortunately, this uncertainty over whether conditions can truly be initial does not prevent us from talking about cosmological fine-tuning.
For most of the history of the universe, many important cosmological quantities are well-described by classical dynamics.
This includes the expansion of the scale factor, as well as the evolution of perturbations, considered as modes in Fourier space of fixed comoving wavelength ({\it i.e.}, expanding along with the universe).%
\footnote{Inflation is an important exception.
During inflation itself, the state of the universe has essentially only one branch.
When inflation ends, reheating creates a large number of excited degrees of freedom, effectively ``measuring'' the state of the inflaton and causing the wave function to split into many branches \cite{Boddy:2014eba}.
This process explains how a perturbed post-inflationary universe can develop out of an unperturbed inflationary state.}
On small scales the dynamics are nonlinear and entropy-generating, due to a variety of processes such as star formation, supernovae, magnetic fields, and violent gravitational relaxation.
Consequently, the evolution at those wavelengths is not well-approximated by reversible equations on phase space.
This leaves us, however, with the dynamics on large scales -- in the present universe, more then ten million light-years across -- that can be treated as the Hamiltonian evolution of an autonomous set of degrees of freedom.
Therefore, we can circumvent conceptual problems raised by the idea of ``initial conditions'' by simply asking whether the trajectory of the large-scale universe since the aftermath of the Big Bang is natural, or fine-tuned, in the space of all such trajectories.

%%%%%%%%%%%%%%%%%%%%%%%%%%%%%%%%%%%%%%%%%%%%%
\subsection{The Canonical Measure}

Fortunately, it is possible to construct a preferred measure on the space of trajectories, which we can use to judge the amount of fine-tuning exhibited by our real universe.
We start by considering the measure on phase space itself, and use that to find a measure on the space of paths through phase space that represent solutions to the equations of motion.%
\footnote{This notion of a cosmological measure is completely separate from that which arises in what is sometimes called ``the measure problem in cosmology, which deals with the relative frequency of different kinds of observers in a multiverse.
See {\it e.g.} \cite{Winitzki:2006rn,Aguirre:2006ak,Linde:2008xf,SchwartzPerlov:2010ne,Freivogel:2011eg,Salem:2011qz}.}

In classical mechanics, there is a natural measure on phase space $\Gamma$, the Liouville measure.
To construct it in terms of coordinates $\{q^i\}$ and momenta $\{p_i\}$, we first write down the symplectic $2$-form on $\Gamma$,
\begin{equation}
\omega=\sum_{i=1}^{n}\mathrm{d}p_{i}\wedge\mathrm{d}q^{i}. 
\label{symplectic}
\end{equation}
Note that the dimension of $\Gamma$ is $2n$.
The Liouville measure is then a $2n$-form given by
\begin{equation}
\Omega=\frac{\left(-1\right)^{n\left(n-1\right)/2}}{n!}\omega^{n}.
\label{liouville}
\end{equation}
All that matters for our current purposes is that such a measure exists, and is uniquely defined.
What makes the Liouville measure special is that it is conserved under time evolution.
That is, given states that initially cover some region $S\subset\Gamma$ and that evolve under Hamilton's equations to cover region $S^{\prime}$, we have 
\begin{equation}
\int_{S}\Omega=\int_{S^{\prime}}\Omega.
\end{equation}

Classical statistical mechanics assumes that systems in equilibrium have a probability distribution in phase space that is uniform with respect to this measure, subject to appropriate macroscopic constraints.
Meanwhile, in connecting cosmology with statistical mechanics, we assume that the microstate of the early universe is chosen randomly from a uniform distribution in the Liouville measure, subject to the (severe) constraint that the macrocondition has the low-entropy form given by the Past Hypothesis.
Albert \cite{albert} calls this the ``Statistical Postulate".%
\footnote{Albert and Loewer have referred to the combined package of the dynamical laws, the Past Hypothesis, and the Statistical Postulate as the ``Mentaculus," from the Coen brothers film {\it A Serious Man} (see {\it e.g.} \cite{loewer2012emergence}).}
Of course one can question why probabilities should be uniform in \emph{this} measure rather than some other one.
Even if the Liouville measure is somehow picked out by the dynamics by virtue of being conserved under evolution, we are nevertheless free to construct any measure we like.
For our present purposes, this kind of question seems misguided.
As discussed in Section~\ref{what}, the point of fine-tuning arguments is to find clues that can guide us to inventing more comprehensive physical theories.
We are not arguing for some metaphysical principle to the effect that the universe \emph{should} be chosen uniformly in phase space according to the Liouville measure; merely that, given this measure's unique status as being picked out by the dynamics, states that look natural in this measure tell us very little, while states that look unnatural might reveal useful information.
(See \cite{Schiffrin:2012zf} for a critique of the use of cosmological measures in the way I am advocating here.)

In general, having a measure on phase space $\Gamma$ does not induce a natural measure on the space of trajectories $T$, which is one dimension lower.
(There is a natural map $\Gamma \rightarrow T$, which simply sends each point to the trajectory it is on; however, while differential forms can be pulled back under such maps, they cannot in general be pushed forward \cite{Carroll:2004st}.)
In the case of general relativity, Gibbons, Hawking and Stewart (GHS, \cite{Gibbons:1986xk}) showed that there is nevertheless a unique measure satisfying a small number of reasonable constraints: it is positive, independent of coordinate choices, respects the symmetries of the theory, and does not require the introduction of any additional structures.
GHS relied on the fact that general relativity is a constrained Hamiltonian system: because the metric component $g_{00}$ is not a propagating degree of freedom in the Einstein--Hilbert action, physical trajectories obey a constraint of the form $\mathcal{H}=\mathcal{H}_{\star}$, where $\mathcal{H}$ is the Hamiltonian and $\mathcal{H}_{\star}$ is a constant defining the constraint surface.
(In cosmological spacetimes we generally have $\mathcal{H}_{\star}=0$.)
The space $U$ of physical trajectories -- those obeying the Hamiltonian constraint -- is thus two dimensions lower than the full phase space $\Gamma$.
GHS construct a measure on $U$ by identifying the $n$th coordinate on phase space as time $t$, for which $\mathcal{H}$ is the conjugate variable.
The symplectic form (\ref{symplectic}) is then
\begin{equation}
\omega= \tilde{\omega}+\mathrm{d}\mathcal{H}\wedge\mathrm{d}t,
\end{equation}
where
\be
  \tilde{\omega}\equiv \sum_{i=1}^{n-1}\mathrm{d}p_{i}\wedge\mathrm{d}q^{i}.
\ee
GHS show that the $\left(2n-2\right)$-form
\begin{equation}
\Theta=\frac{\left(-1\right)^{\left(n-1\right)\left(n-2\right)/2}}{\left(n-1\right)!}\tilde{\omega}^{n-1}
\label{GHSmeasure}
\end{equation}
is a unique measure satisfying their criteria.

The GHS measure (\ref{GHSmeasure}) has the attractive feature that it is expressed locally in phase space.
Therefore, it can be evaluated on the space of trajectories simply by choosing some transverse surface through which all trajectories (or all trajectories of interest for some purpose) pass, and calculating $\Theta$ on that surface; the results are independent of the surface chosen.
In cosmology, for example, we might choose surfaces of constant Hubble parameter, or constant energy density, and evaluate the measure at that time.
This feature has a crucial consequence: the total measure on some particular kind of trajectories, such as ones that are spatially flat or ones that are relatively smooth at some particular cosmological epoch, is completely independent of what the trajectories are doing at some other cosmological epoch.
Therefore, changing the dynamics in the early universe (such as modifying the potential for an inflaton field) cannot possible change the fraction of trajectories with certain specified properties in the late universe.
(Adding additional degrees of freedom can, of course, alter the calculation of the measure.)
New physics cannot change ``unnatural'' trajectories into ``natural'' ones.

At heart, there is not much conceptual difference between studying the purported fine-tuning of the universe in terms of the measure on trajectories and quantifying the low entropy of the early state.
There are relatively few initial conditions with low entropy, and the trajectories that begin from such conditions will have a small measure.
As discussed in Section~\ref{what}, in order to quantify fine-tuning, we generally need to specify a coarse-graining on the space of states as well as a measure.
In the language of trajectories, this corresponds to specifying macroconditions the trajectories must satisfy. 
One benefit of the trajectory formalism is that it is relatively straightforward to ask questions that conditionalize over macroconditions specified at a different time; for example, we can talk about the fraction of the trajectories that are smooth at one time given that they are smooth at some other time.
Another benefit, and a considerable one, is that we can look at features of cosmic evolution that we truly understand without claiming to have full control over the space of states (as would be necessary to completely understand the entropy).
An objection to Penrose's argument is sometimes raised that we don't know enough about how to calculate the entropy in quantum gravity to make any statements at all; using the measure on classical trajectories allows us to make fine-tuning arguments while remaining wholly in a regime where classical general relativity should be completely valid.

%%%%%%%%%%%%%%%%%%%%%%%%%%%%%%%%%%%%%%%%%%%%%
\subsection{Flatness}\label{flatness}

An interesting, and surprisingly nontrivial, application of the GHS measure is to the flatness problem -- as we will see, it doesn't really exist.
(In this section and the next I am drawing on work from \cite{Carroll:2010aj}.)
Consider a Robertson-Walker universe with scale factor $a(t)$ and curvature parameter $\kappa$, obeying the Friedmann equation (\ref{friedmann}), with an energy density from components $\rho_i$ that each scale as a power law, $\rho_i = \rho_{i0} a^{-n_i}$ for some fixed $n_i$. 
This includes the cases of nonrelativistic matter, for which we have $n_M=3$, and radiation, for which $n_R=4$.
Then we can define the corresponding density parameters $\Omega_i$, as well as an ``effective density parameter for curvature,'' via
\be
  \Omega_i\equiv \frac{8\pi G \rho_{i0} a^{-n_i}}{3H^2},\quad
  \Omega_\kappa \equiv -\frac{\kappa}{a^2H^2}.
  \label{densityparameters}
\ee
The Friedmann equation then implies that $\sum_i\Omega_i + \Omega_\kappa = 1$.
The ratio of the curvature to one of the densities evolves with time as
\be
  \frac{\Omega_\kappa}{\Omega_i} \propto a^{n_i-2}.
\ee
Whenever $n_i>2$, as for matter or radiation, the relative of importance of curvature grows with time.
The conventional flatness problem is simply the observation that, since the curvature is not very large today, it must have been extremely small indeed at early times.
Roughly speaking (since details depend on the amounts of matter, radiation, and vacuum energy, as well as their evolutions), we must have had $\Omega_\kappa/\Omega_{\mathrm{matter/radiation}} < 10^{-55}$ in the very early universe in order that the curvature not dominate today.

As we have argued, however, such a statement only has impact if the set of trajectories for which $\Omega_\kappa/\Omega_{\mathrm{matter/radiation}} < 10^{-55}$ in the very early universe is actually small.
It \emph{seems} small, since $10^{-55}$ is a small number.
But that just means that it would be small if trajectories were chosen uniformly in the variable $\Omega_\kappa/\Omega_{\mathrm{matter/radiation}}$, for which we have given no independent justification.
Clearly, this is a job for the GHS measure.

To make things quantitative, consider a Robertson-Walker universe containing a homogeneous scalar field $\phi(t)$ with potential $V(\phi)$.
(A scalar field is more directly applicable than a matter or radiation fluid, since the scalar has a Hamiltonian formulation; one can, however, consider scalar field theories that mimic the behavior of such fluids, so no real generality is lost.)
The scale factor will obey the Friedmann equation (\ref{friedmann}), with the energy density given by
\be
  \rho_\phi = \frac{1}{2}\dot\phi^2 + V(\phi).
\ee
The dynamical coordinates for this model are $a$ and $\phi$, with conjugate momenta $p_a$ and $p_\phi$, as well as a Lagrange multiplier $N$ (the lapse function) that enforces the Hamiltonian constraint.
The lapse function is essentially the square root of the $00$ component of the metric; it is non-dynamical in general relativity, since no time derivatives of $g_{00}$ appear in the action.
Setting $8\pi G=1$ for convenience, the Einstein-Hilbert Lagrangian for the scale factor coupled to the scalar field is 
\begin{equation}
\mathcal{L}=3\left(Na\kappa-\frac{a\dot{a}^{2}}{N}\right)+a^{3}\left[\frac{\dot{\phi}^{2}}{2N}-NV\left(\phi\right)\right].
\end{equation}
The canonical momenta, defined as $p_{i}=\partial\mathcal{L}/\partial\dot{q}^{i}$,
are 
\begin{equation}
p_{N}=0,\qquad p_{a}=-6N^{-1}a\dot{a},\qquad \mathrm{and}\qquad p_{\phi}=N^{-1}a^{3}\dot{\phi}.\label{momenta}
\end{equation}
Performing a Legendre transformation, the Hamiltonian is 
\begin{equation}
\mathcal{H}  =N\left[-\frac{p_{a}^{2}}{12a}+\frac{p_{\phi}^{2}}{2a^{3}}+a^{3}V\left(\phi\right)-3a\kappa\right].
\label{Hamiltonian}
\end{equation}
The equation of motion for $N$ sets it equal to an arbitrary constant, which we can choose to be unity. 
Varying the action with respect to $N$ gives the Hamiltonian constraint, $\mathcal{H}_{\star}=0$, which is equivalent to the Friedmann equation.
Setting $N=1$, the remaining phase space $\Gamma$ is four-dimensional.
The Hamiltonian constraint surface is three-dimensional, and the space of physical trajectories $U$ is two-dimensional. 
The GHS measure can be written as the Liouville measure subject to the Hamiltonian constraint: 
\begin{equation}
\Theta=\left.\left(\mathrm{d}p_{a}\wedge\mathrm{d}a+\mathrm{d}p_{\phi}\wedge\mathrm{d}\phi\right)\right|_{\mathcal{H}=0}.
\label{GHS-RW}
\end{equation}

The measure can be evaluated in expanding Friedmann-Robertson-Walker universes by integrating (\ref{GHS-RW}) over a specified transverse surface, for example by setting $H=H_*$.
The answer works out to be
\begin{align}
  \mu &\equiv  \int_{H=H_*}\Theta \\
 &=  \int_{H=H_*}\Theta_{a\phi}\, da\,d\phi \\
%  &= \sqrt{|3\kappa|^3}  \int_{H=H_*}\frac{6H_*^2 + 2 \dot\phi^2 -2V(\phi)}{|6H_*^2 - \dot\phi^2 -2V(\phi)|^{5/2}} d\phi\, d\dot\phi \\
  &= -6\int_{H=H_*}
   \frac{3a^{3}H_*^2- a^3V + 2ak}
  {\left(6a^{2}H_*^2- 2a^2V+ 6 k\right)^{1/2}}\,da\, d\phi .
  \label{ghs3}
\end{align}
We can make this expression look more physically transparent by introducing the variable
\be
  \Omega_V\equiv \frac{V(\phi)}{3H^2},
\ee
as well as the curvature density parameter defined in (\ref{densityparameters}).
The scale factor is strictly positive, so that integrating over all values of $\Omega_\kappa$ is equivalent to integrating over all values of $a$.  
The measure is then
\be
  \mu = 3\sqrt{\frac{3}{2}}H_*^{-2}\int_{H=H_*} 
  \frac{1-\Omega_V - \frac{2}{3}\Omega_\kappa}
  {|\Omega_\kappa|^{5/2}\left(1-\Omega_V  -\Omega_\kappa\right)^{1/2}}\,d\Omega_\kappa\, d\phi. 
  \label{flatness1}
\ee

This integral is divergent; it blows up as $\Omega_\kappa \rightarrow 0$, since the denominator includes a factor of $|\Omega_\kappa|^{5/2}$.   
By itself, the divergence isn't surprising; the set of classical trajectories is non-compact.
The more interesting fact is \emph{where} it diverges -- for universes that are spatially flat ($\Omega_\kappa=0$), which is certainly a physically relevant region of parameter space.

This divergence was noted in the original GHS paper \cite{Gibbons:1986xk}, where it was attributed to ``universes with
very large scale factors'' due to a different choice of variables.  
That characterization isn't very useful, since ``large scale factor'' is a feature along the trajectory of any open universe, rather than picking out a particular type of trajectory.
Later works  \cite{Hawking:1987bi,Coule:1994gd,Gibbons:2006pa} correctly described the divergence as arising from nearly-flat universes.  
Gibbons and Turok \cite{Gibbons:2006pa} advocated dealing with the infinity by discarding all flat universes by fiat, and concentrating on the non-flat universes.
Tam and I \cite{Carroll:2010aj} took the opposite view: what (\ref{flatness1}) is telling us is that almost every Robertson-Walker cosmology is spatially flat.
Rather than throwing such trajectories away, we should throw all of the others away and deal with flat universes.

What one wants, therefore, is a measure purely on the space of flat universes.
The procedure we advocated in \cite{Carroll:2010aj} for obtaining such a measure was faulty, as our suggested regularization gave a result that was not invariant under a choice of surface on which to evaluate the measure.
This problem was later solved in \cite{Remmen:2013eja}, which derived a measure on the phase space for flat universes by demanding that it obey Liouville's theorem, and \cite{Remmen:2014mia}, which used this to derive a measure on the space of trajectories.
Applying this to the case of inflation, we showed that one generically expects a very large amount of inflation for unbounded potentials such as $V(\phi) \propto \phi^2$, and relatively few $e$-folds for ``natural'' inflation \cite{Freese:1990rb} in which $V(\phi) \propto \cos(\phi)$.

From the point of view of fine-tuning, using the GHS measure completely alters our picture of the flatness problem.
We noted that the conventional formulation of the problem implicitly assumes a measure that is uniform in $\Omega_\kappa$, which seemed intuitively reasonable.
But in fact the measure in the vicinity of flat universes turns out to be proportional to $1/|\Omega_\kappa|^{5/2}$, which is a dramatic difference.
Rather than sufficiently flat universes being rare, they are actually generic.
We take this result to indicate that the flatness problem really isn't a problem at all; it was simply a mistake, brought about by considering an informal measure rather than one derived from the dynamics.

%%%%%%%%%%%%%%%%%%%%%%%%%%%%%%%%%%%%%%%%%%%%%
\subsection{Smoothness}

The surprising result that almost all universes are spatially flat might raise the hope that a careful consideration of the measure might also explain the smoothness of the universe: perhaps almost all cosmological trajectories are extremely smooth at early times.
Sadly, the opposite is true, as can be seen by extending the GHS measure to perturbed spacetimes \cite{Carroll:2010aj}.
This might seem like a difficult task, since there are many ways the universe can be perturbed.
But as long as the perturbations are small, every Fourier mode evolves independently according to a linear equation of motion.
Therefore, we can consider the measure on a mode-by-mode basis.

For linear scalar perturbations, the coupled dynamics of a matter/radiation fluid and the spacetime curvature in a background spacetime can be described by a single degree of freedom, the cosmological perturbation field $u(\vec{x},t)$, as discussed by Mukhanov, Feldman and Brandenberger \cite{Mukhanov:1990me}.
Given the action for this field, we can isolate the dynamical variables and construct  the symplectic two-form on phase space, which can then be used to compute the measure on the  set of solutions to Einstein's equations.  
Various subtleties arise along the way, but the final answer is relatively straightforward.
Here I will just quote the results; calculations can be found in \cite{Carroll:2010aj}.

It is convenient to switch to conformal time,
\be
  \eta = \int a^{-1}dt.
\ee
Derivatives with respect to $\eta$ are denoted by the superscript $'$, and $\htilde \equiv a'/a$ is related to the Hubble parameter $H=\dot{a}/a$ by $\htilde = aH$. 
In Fourier space the cosmological perturbation field is a function $u(\vec{k}, t)$, where $\vec{k}$ is the comoving wave vector.
It is essentially a scaled version of the $00$ component of the metric perturbation, which is just the Newtonian gravitational potential:
\be
  \Phi = (\bar{\rho}+\bar{p})^{1/2}u,
\ee
where $\bar\rho$ and $\bar{p}$ are the background energy density and pressure.
From that we can express the energy density perturbation in conformal Newtonian gauge,
\be
    \delta\rho = \frac{1}{4\pi G a^2}\left[\nabla^2\Phi - 3\htilde(\Phi' + \htilde\Phi)\right].
\ee

The cosmological perturbation field obeys an equation of motion
\be
  u'' -c_s^2\nabla^2u - \frac{\theta''}{\theta} u = 0.
  \label{ueq}
\ee
Here, $c_s$ is the speed of sound in the matter/radiation fluid, and $\theta(\eta)$ is a time-dependent parameter given by
\be
  \theta =\frac{1}{a}\left[\frac{2}{3}\left(1 - \frac{\htilde'}{\htilde^2}\right)\right]^{-1/2}.
\ee
This equation of motion can be derived from an action
\be
  S_u = \frac{1}{2}\int d^4x \left(u'^2 - c_s^2 \sum_i\partial_iu \partial_iu + \frac{\theta''}{\theta}u^2 \right).
\ee
Defining the conjugate momentum $p_u = \partial{\mathcal L}/\partial u' = u'$, we can describe the
dynamics in terms of a Hamiltonian for an individual mode with wavenumber $k$, 
\begin{equation}
  \mathcal{H} = \frac{1}{2}p_u^2 + \frac{1}{2}\left(c_s^2k^2  - \frac{\theta''}{\theta}\right)u^2.
  \label{pertham}
\end{equation}
This is simply the Hamiltonian for a single degree of freedom with a time-dependent
effective mass $m^2= c_s^2k^2  - \theta''/\theta$.

One convenient hypersurface in which we can evaluate the flux of trajectories is $\eta=
\eta_*=\mbox{constant}$.  
A straightforward calculation shows that the measure evaluated on such a surface is simply
\be
  \mu = \int_{\eta=\eta_*} \,du\,dp_u.
  \label{pertmeasure}
\ee
In other words, the measure on a perturbation mode is completely uniform in the $\{u, p_u\}$ variables, much as we might have na\"ively guessed, and in stark contrast to the flatness problem.
All values for $u$ and $p_u$ are equally likely; there is nothing in the measure that would explain the small 
observed values of perturbations at early times.  
Hence, the observed homogeneity of our universe does imply considerable fine-tuning.

We can use this measure to roughly quantify how much fine-tuning is involved in the conventional assumption of a smooth universe near the Big Bang.
For purposes of convenience, \cite{Carroll:2010aj} asked a simple question: assuming that the universe had the observed amount of uniformity at the surface of last scattering ($\delta\rho/\bar{\rho}\leq 10^{-5}$), what fraction of the allowed trajectories were also smooth in the very early universe, say near the scale of grand unification when the energy density was $\rho \sim (10^{16}\,\mathrm{GeV})^4$?
The answer is, unsurprisingly, quite small.
For each individual mode, the chance that it was small at the GUT scale given that it is small at recombination is of order $10^{-66}$.
But there are many modes, and if any one of them is large then the universe is not truly smooth. 
The total fraction of universes that were smooth at early times is just the product of the fractions corresponding to each mode.
Choosing reasonable bounds for the largest and smallest modes considered, the total fraction of trajectories that are smooth at early times works out to be
\be
  f(\textrm{smooth at GUT scale}|\textrm{smooth at recombination}) \approx 10^{-6.6\times 10^7}.
  \label{fraction}
\ee
This represents a very conservative estimate for the amount of fine-tuning involved in the standard cosmological model.

It might seem strange to ask a conditional question that assumes the universe was smooth at the time of last scattering, rather than directly inquiring about the fraction of universes that were smooth at recombination.
But that fraction is ill-defined, since the phase space of perturbations is unbounded.
We could have calculated the fraction of universes that are relatively smooth today that were also smooth at recombination, and obtained a similarly tiny number.
More importantly, we are interested in the fine-tuning necessary for the universe to obey the Past Hypothesis.
The reason why (\ref{fraction}) is such a small number is that most trajectories that are smooth at last scattering contain modes that were large at earlier times but decayed.
That is morally equivalent to trajectories that start with relatively high entropy, but that start with delicate correlations that cause the entropy to decrease as time passes.
All of conventional cosmology assumes that the early universe was not like that; (\ref{fraction}) quantifies the amount of fine-tuning implied by this assumption.

We can therefore conclude that the smoothness of the early universe does indeed represent an enormous amount of fine-tuning.
In reaching this conclusion, we made no reference to the causal structure nor to any thwarted attempts at equilibration.
Those considerations, which play a central role in formulating the horizon problem, are red herrings.
The real sense in which the early universe was fine-tuned is extremely simple: the overwhelming majority of cosmological trajectories, as quantified by the canonical measure, are highly nonuniform at early times, and we don't think the real universe was like that.
Clearly, the specific numerical value we obtain is not of central importance; what is certain is that the history of our actual universe does not look anything like it was chosen randomly.

%%%%%%%%%%%%%%%%%%%%%%%%%%%%%%%%%%%%%%%%%%%%%
\section{Discussion}\label{discussion}

I have argued that the traditional discussion of the fine-tuning of the early universe in terms of the horizon and flatness problems is misguided.
The flatness problem is based on an implicit use of an unjustified measure on the space of initial conditions; when a more natural measure is used, we see that almost all Robertson-Walker universes are spatially flat.
The difficulty with the horizon problem is not that it is incorrect, but that it is inconclusive, since we don't have a clear picture of how the situation would change if distant regions on the surface of last scattering actually had been in causal contact.
It is much more sensible to quantify fine-tuning in terms of the measure on the space of cosmological trajectories.
From that perspective, it is clear that the overwhelming majority of such trajectories, conditionalized on some reasonable requirement in the late universe, are wildly inhomogeneous at early times.
The fact that the actual early universe was not like that, as specified by the Past Hypothesis, gives us a clear handle on the kind of fine-tuning we are faced with.

As with most discussions of fine-tuning, in this paper I have prejudiced the discussion somewhat by assuming throughout that the universe is an approximately Robertson-Walker cosmology with a certain (large) amount of matter and radiation.
While we have seen that the universe is fine-tuned even given that framework, such an assumption is much stronger than what is required, for example, by the anthropic principle.
Life can certainly exist in universes with far fewer stars and galaxies than what we observe.
In the presence of a stable vacuum energy, the highest-entropy configuration for the universe to be in is empty de~Sitter space \cite{Carroll:2004pn}.
The worry there is that vacuum fluctuations give rise to an ensemble of freak ``Boltzmann brain'' observers \cite{Dyson:2002pf,Albrecht:2004ke,Bousso:2006xc}.%
\footnote{The problem is not that a de~Sitter scenario predicts that ``we should be Boltzmann brains."
Rather, we should consider ourselves to be one of any of the many observers that find themselves in precisely our current cognitive situation.
For most such observers, that cognitive situation -- {\it e.g.}, my current belief that there is a person named ``David Albert'' who wrote a book entitled {\it Time and Chance} -- is completely uncorrelated with the reality of their actual environment.
Such a situation is cognitively unstable, for reasons explained in related contexts in \cite{albert} (at least, to the best of my current recollection).}
As argued in \cite{Boddy:2014eba}, however, quantum fluctuations in de~Sitter space don't actually bring into existence decohered branches of the wave function containing such freak observers.
Nevertheless, it seems reasonable to think that the space of trajectories containing one person or one galaxy in an otherwise empty background has a much greater measure than the kind of universe in which we live, with over a hundred billion galaxies; at least, such a situation has a much higher entropy.
We are therefore still left with the fundamental cosmological question: ``Why don't we live in a nearly-empty de~Sitter space?''

Formulating cosmological fine-tuning in the language of the measure on trajectories puts the inflationary-universe scenario in its proper context.
A major original motivation of inflation was to solve the horizon and flatness problems.
Reformulating the issue raised by the horizon problem as a matter of the measure on cosmological trajectories brings the problem with inflation into sharp focus: the fact that most trajectories that are homogeneous at late times are highly non-homogeneous at early times is completely independent of physical processes in the early universe.
It depends only on the measure evaluated at relatively late times.
Inflation, therefore, cannot solve this problem all by itself.
Indeed, the measure reinforces the argument made by Penrose, that the initial conditions necessary for getting inflation to start are extremely fine-tuned, more so than those of the conventional Big Bang model it was meant to help fix.
Inflation does, however, still have very attractive features.
It posits an initial condition that, while very low-entropy, is also extremely simple, not to mention physically quite small.
(With inflation, our observable universe could have been one Planck length across at the Planck density; without inflation, the same patch was of order one centimeter across at that time.
That is an incredibly large volume, when considered in Planck units, over which to have initial homogeneity.)
Therefore, while inflation does not remove the need for a theory of initial conditions, it gives those trying to construct such a theory a relatively reasonable target to shoot for.

Of course, all of this discussion about fine-tuning and the cosmological measure would be completely pointless if we did have a well-formulated theory of initial conditions (or, better, of our cosmological history considered as a whole).
Ultimately the goal is not to explain why our universe appears unnatural; it's to explain why we live in this specific universe.
Making its apparent unnaturalness precise is hopefully a step toward achieving this lofty ambition.

%%%%%%%%%%%%%%%%
\section*{Acknowledgments}

It is a pleasure to thank David Albert for inspiration and conversations over the years, Barry Loewer for his patience, Shelly Goldstein for useful discussions, my collaborators Heywood Tam and Grant Remmen for their invaluable insights, and Tim Maudlin for the interactions that proximately inspired this paper.
This research is funded in part by DOE grant DE-SC0011632, and by the Gordon and Betty Moore Foundation through Grant 776 to the Caltech Moore Center for Theoretical Cosmology and Physics.
%%%%%%%%%%%%%%%%%%%%%%%%%%%%%%%%%%%%%%%%%%%%%%%%%%%%%

%\bibliographystyle{ieeetr}
%\bibliographystyle{report}
\bibliographystyle{utphys}

\bibliography{arrow-bib}
\end{document}